\documentclass[aps,pre,twocolumn,superscriptaddress,groupedaddress]{revtex4}  
\usepackage{graphicx}  
\usepackage{dcolumn}   
  
\usepackage{bm}        
\usepackage{amssymb}   
\usepackage{mathtools}
\usepackage{amsmath,amsthm,amssymb,amsthm,dsfont,mathrsfs,bbm}
\usepackage{braket}
\usepackage[T1]{fontenc} 
\usepackage{newtxtext,newtxmath} 

\usepackage[usenames,dvipsnames]{xcolor}
\usepackage{tikz,pgfplots,tkz-berge,tkz-graph,bbm,bm,mathrsfs,dsfont}

\usepackage[font=small,labelfont=bf]{caption}
\usepackage{hyperref}
\hypersetup{
 colorlinks=true}
\hypersetup{pdfauthor={Malatesta Parisi Rizzo},pdftitle={Two-Loop Corrections to Large Order Behavior of pi^4 Theory},%
 colorlinks, linktocpage=true, pdfstartpage=3, pdfstartview=FitV,%
 breaklinks=true, pdfpagemode=UseNone, pageanchor=true, pdfpagemode=UseOutlines,%
 urlcolor=orange, linkcolor=blue, citecolor=red
}
\usepackage{feynmp}
\ifpdf
\else
\fi

\newcommand\manubrio[1] {\;\,\parbox{#1}{\fmfreuse{manubrio} } }
\newcommand\sunset[1] {\;\,\parbox{#1}{\fmfreuse{sunset} } }
\newcommand\eight[1] {\;\,\parbox{#1}{\fmfreuse{eight} } }
\newcommand\tadpole[1] {\;\,\parbox{#1}{\fmfreuse{tadpole} } }
\newcommand\plainx[1] {\;\,\parbox{#1}{\fmfreuse{plainx} } }
\newcommand\blobx[1] {\;\,\parbox{#1}{\fmfreuse{blobx} } }
\newcommand\tadpoleJac[1] {\;\,\parbox{#1}{\fmfreuse{tadpoleJac} } }
\newcommand\plainxJac[1] {\;\,\parbox{#1}{\fmfreuse{plainxJac} } }
\newcommand\plainxJacTwo[1] {\;\,\parbox{#1}{\fmfreuse{plainxJacTwo} } }

\newcommand \gatto{Schr\"odinger }

\begin{document}
 

 
 \title{Two-Loop Corrections to Large Order Behavior of $\varphi^4$ Theory}
 \author{Enrico M. Malatesta}
 \affiliation{
  Universit\`a di Milano, Dip.~di Fisica and INFN, Sezione di Milano, Via Celoria 16,  I-20133 Milan, Italy
 }

 \author{Giorgio Parisi}
 \affiliation{Dip. Fisica, Universit\`a ``Sapienza", Piazzale A. Moro 2, I-00185, Rome, Italy}
 \affiliation{INFN, Sezione di Roma I,  CNR-NANOTEC UOS Roma}
 
 \author{Tommaso Rizzo}
 \affiliation{Dip. Fisica, Universit\`a ``Sapienza", Piazzale A. Moro 2, I-00185, Rome, Italy}
 \affiliation{ISC-CNR, UOS Rome, Universit\`a ``Sapienza'', Piazzale A. Moro 2, I-00185, Rome, Italy}


\begin{abstract}
 We consider the large order behavior of the perturbative expansion of the scalar $\varphi^4$ field theory in terms of a perturbative expansion around an instanton solution. We have computed the series of the free energy up to two-loop order in two and three dimension. Topologically there is only an additional Feynman diagram with respect to the previously known one dimensional case, but a careful treatment of renormalization is needed. The propagator and Feynman diagrams were expressed in a form suitable for numerical evaluation. We then obtained explicit expressions summing over $O(10^3)$ distinct eigenvalues determined numerically with the corresponding eigenfunctions.   
\end{abstract}
 
\maketitle

\section{Introduction}
Perturbation theory is one of the few powerful analytical tools through which one can calculate physical quantities in an interacting field theory \cite{ZJ}. However, as pointed out for the first time by Dyson in the framework of QED \cite{Dyson}, the most common perturbation expansions we use in field theory are not convergent, but only asymptotic. In addition the expansion parameter may not be always small: this is the case, unlike QED, of critical phenomena \cite{ParisiLibro}. It is therefore essential to understand how to extract numerical and physical information from the perturbative coefficients of the series. One of the first attempts to sum the perturbative series were given by \cite{OnlyPadè}, using Pad\`e approximants. However, the convergence couldn't be tested since they used only a few perturbative coefficients. Then it was proposed to sum the perturbative series using the Borel summation method, combined to Pad\`e approximants \cite{Baker&Nickel}. It was subsequently realized that, when one knows the behavior of the coefficients of the series at large orders, one can sum the series using a much more powerful method that combines a Borel transformation and a conformal mapping. This has allowed to estimate critical exponents with a very good precision and for a large variety of models \cite{CrExp1,CrExp2,CrExp3,CrExp4,CrExp5,KondoModel,GuidaZinn}. The behavior of the perturbative series at large order has been of extreme importance not only in understanding the domain of validity of the theory, but also because we can use this information to accurately determine physical quantities, even if the coupling constant is not so small. Bender and Wu \cite{Bender&Wu1, Bender&Wu2}, using WKB method, succeeded in evaluating the large order behavior of the one-dimensional quartic anharmonic oscillator, showing how it is related to barrier penetration effects when the sign of the coupling constant is negative. Then Lipatov and others \cite{Lam, Lipatov1,Lipatov2,Brezin&Guillou&ZJ}, using a dispersion relation approach, proved that the large order behavior of perturbation theory is related to the instability of the theory for negative values of the coupling constant and that it can then be obtained using the so-called ``pseudoparticle'' or \emph{instanton} solutions of the classical equation of motion. This method \cite{Jen&Zinn} not only gives the same results of semiclassical calculations concerning the one-dimensional quantum anharmonic oscillator \cite{Auberson,Collins&Soper}, but can also be extended to the field theory case and can in principle be used for more complex types of interactions. In this paper we will focus on the $\varphi^4$ field theory in $d<4$, where the Borel summability of the perturbative series has been rigorously proven \cite{Loeffel,Simon,Eckmann}. Expanding around the instanton saddle point adds some technical difficulties since it  breaks explicitly the time-translational invariance of the theory. This induces zero modes over which one must integrate via the \emph{collective coordinates} method \cite{Zittartz}. In 2 and 3 dimension, one has another, additional problem: renormalization. Order by order the mass counterterm that makes the original perturbative series finite, should also renormalize the instanton perturbative series. At one loop order, the only divergence occurs in the calculation of the determinant of the fluctuation operator orthogonal to zero modes and indeed this is canceled by the first order expansion of the mass counterterm \cite{Parisi}. The two-loop order series of the partition function has been computed in the $d=1$ case (quantum mechanics) both with WKB and Lipatov method \cite{Auberson,Collins&Soper, Jen&Zinn}. In that case, since one can write down the explicit formula for the propagator, it is also possible to analytically evaluate the correction. 
In this work we compute the two-loop order series of the partition function in $d=2$ and $d=3$. 

The paper is organized as follows. In Section \ref{Section2} we will review the one loop result, for completeness. In Section \ref{Section3} we will perform the expansion to the two loop order of the action. In Section \ref{Section4} the two loop correction to the large order behavior of the partition function is written in terms of diagrams, and their renormalization is discussed. We will see also how one last divergence in $Z_k$, which is not completely canceled by the counterterms, is eliminated once we evaluate the large order behavior of the free energy. In Section \ref{Section5} we will express the diagrams in a form suitable to numerical integration and give our results.
In Section \ref{Conclusions} we present our conclusions. 


\section{One Loop Computation} \label{Section2}  
\begin{figure}[!ht]
	\centering 
	\includegraphics[ ]{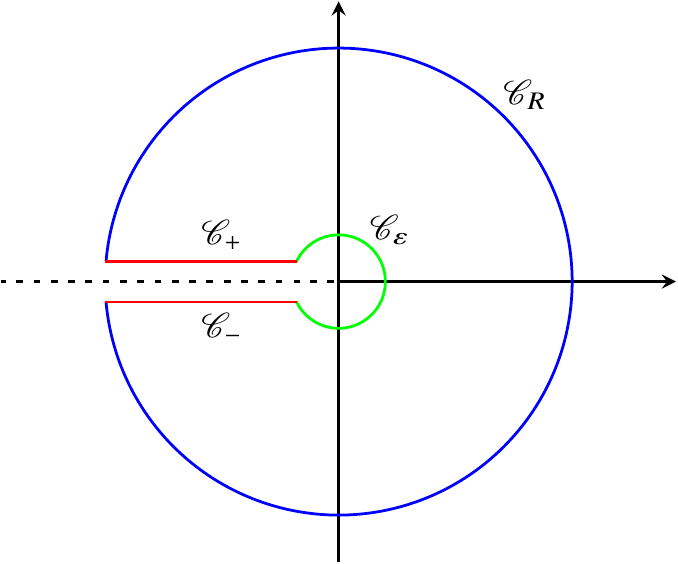} 
	\caption{Contours used to derive the dispersion relation. The dashed line represents the branch cut at $(-\infty,0)$.}
	\label{fig:Contour}
\end{figure}

The $\varphi^4$ partition function has the following expression 
\begin{subequations}
	\begin{align}
	& Z(g) = \int \mathscr{D} \varphi \; e^{-S(\varphi)} = \sum_{k=0}^{\infty} Z_k \, g^k \;, \\
	& S(\varphi) = \int d^d x \left[ \frac{1}{2} \left( \nabla \varphi \right)^2 + \frac{1}{2} \varphi^2 + \frac{g}{4} \varphi^4 + \frac{\delta m^2}{2} \varphi^2  \right]
	\end{align} 
\end{subequations} 
where $\delta m^2$ is the mass counterterm. We normalize the path integral with respect to the free partition function $Z(0)$. $Z(g)$ is an analytic function in the whole complex plane, with a cut on the real negative axis. Using Cauchy integral formula we have 
\begin{equation} 
Z_k = \frac{1}{2\pi i} \int_{\mathscr{C}}  dg \, \frac{ Z(g)}{g^{k+1}} \,.
\end{equation}
where $\mathscr{C}$ is the path in the complex plane represented in Fig. \ref{fig:Contour}.  
Since $g Z(g) \to 0$ for $g\to 0$ and $Z(g)$ decays polynomially to zero for $g \to \infty$, the contribution from paths $\mathscr{C}_\varepsilon$ and $\mathscr{C}_R$ vanishes in the limit $\varepsilon \to 0$ and $R \to \infty$ respectively. As $\mathscr{C}_+$ and $\mathscr{C}_-$ are complex conjugate paths we can write down a relation that connects the perturbative coefficient $Z_k$ to the imaginary part of $Z(g)$   
\begin{equation}
\label{DispersionRelation}
Z_k = \frac{1}{\pi} \int_{-\infty}^{0} dg \, \frac{\text{Im} \, Z(g)}{g^{k+1}} \,.
\end{equation}
In order to evaluate $Z_k$ for $k \gg 1$, we have first to estimate $\text{Im} \, Z(g)$ for $g$ small and \emph{negative}.
The standard way to do so, is to analytically continue $Z(g + i \varepsilon)$ and $Z(g - i \varepsilon)$ rotating clockwise and anticlockwise the integration contour in the $\varphi$ functional space as $\arg g \to \pm \pi$. 
However, in order to set-up the perturbation expansion of $\text{Im} \, Z$,  we will formally use the representation \cite{ZJ}
\begin{equation}
\label{ImZ}
\text{Im} \, Z(g) = \frac{1}{2i} \int \! \mathscr{D} \varphi \; e^{-S(\varphi)} 
\end{equation}
where the integral is performed over the real axis. One should remember that this expression is formal because the integral is not convergent: since $\text{Im} \, Z(g)$ is real, we expect that the determinant of fluctuation around the saddle point will be negative.  
$\text{Im} \, Z(g)$ is dominated by solutions of the equation of motion
\begin{equation}
\label{EOM}
\left( - \Delta  + 1 \right) \varphi_c(x) + g \varphi^3_c(x) = 0 \,, 
\end{equation}
that depend explicitly on the $x$ space-time variable and that are non-vanishing in the large volume limit \cite{ZJ}. These solutions are called \emph{instantons}. The mass counterterm is sub-leading because the saddle point solution is of order $(-g)^{-1/2}$. Rescaling 
\begin{equation}
\label{InstantonScaling}
\varphi_c(x) = \frac{\Phi_c(x)}{\sqrt{-g}} \,,
\end{equation}
we obtain an equation for $\Phi_c(x)$ independent of $g$
\begin{equation}
\label{EOM2}
\left( - \Delta  + 1 \right) \Phi_c(x) - \Phi^3_c(x) = 0 \,, 
\end{equation}
The minimum of the action is reached by a spherically symmetric solution \cite{ZJ, Coleman&Glazer}, i.e. $\Phi_c(x) = \Phi_c(\left| x \right|) = \Phi_c( r ) $. This solution is infinitely degenerate because if $\Phi_c( \left| x \right| )$ satisfies (\ref{EOM2}), also $\Phi_c( \left| x -x_0\right| )$ with $x_0 \in \mathds{R}^d$ is a solution. Therefore choosing an arbitrary value of the origin $x_0$ and setting $r = \left|x-x_0\right|$, we can write 
\begin{equation}
\left[ -\frac{d^2}{d r^2}  - \frac{d-1}{r} \frac{d}{dr} + 1\right] \Phi_c(r) - \Phi_c^3(r) = 0
\end{equation}
with the boundary condition $\Phi_c(r) \to 0$ for $r \to \infty$. The minimal solution is even in $r$; the initial condition can be imposed with the value of $\Phi_c(r)$ at the origin. Because of the sign ambiguity of this initial condition, we will have two families of solutions; when we sum over those families we will have a factor 2 to take into account. Moreover $\Phi_c(r)$ has no nodes, and decays exponentially to 0 for $r \to \infty$. Only in $d=1$ we can solve analytically this equation and the solution is
\begin{equation}
\label{Instanton1d}
\Phi_c(x) = \frac{\sqrt{2}}{\cosh x} \,, \quad d=1 \,.
\end{equation}
Defining the quantities
\begin{equation}
\label{Integrals}
I_k = \int d^d x \; \Phi^k_c(x) \,,
\end{equation}
we can write the action evaluated on the saddle point as
\begin{equation}
S_c \equiv S(\varphi_c) = -\frac{A}{g} \,,
\end{equation}
where
\begin{equation}
A = \frac{I_4}{4} = \frac{1}{d} \int d^d x \left( \nabla \Phi_c \right)^2 =\frac{I_4-I_2}{d} = \frac{I_2}{4-d} > 0 \,.
\end{equation}
These equalities are referred to as the \emph{Virial Theorem} \cite{ZJ}. In Table \ref{tab:Integrals} we report some numerical estimates of the quantities $I_k$. Next we have to perform the second order functional derivative of the action in order to evaluate Gaussian fluctuation around the instanton saddle point
\begin{equation}
\begin{aligned}
M(x_1,x_2) & = \left. \frac{\delta^2 S}{\delta \varphi(x_1) \delta \varphi (x_2)} \right|_{\varphi = \varphi_c} \\ 
&= \left[ -\Delta_{x_1} + 1 - 3 \Phi_c^2(x_1) \right] \delta(x_1-x_2) \,.
\end{aligned}
\end{equation}
\begin{table}
	\begin{center}
		\begin{tabular}[b]{c|cccc}
			\hline
			$d$ & $I_2$ & $I_3$ &$I_4$&$I_6$ \\ 
			\hline
			1 & 4  & $\sqrt{2} \pi$ &$\frac{16}{3}$& $\frac{128}{15}$\\ 
			2 & 11.700896 & 15.109670 & 23.40179 & 71.080173 \\
			3 & 18.897251 &31.691522 & 75.589005 & 659.868351\\
			\hline
		\end{tabular}
	\end{center}
	\caption{Numerical values of the integrals $I_k$ defined in (\ref{Integrals}) for some $k$ and in $d=2,3$. One dimensional values are analytical.}
	\label{tab:Integrals} 
\end{table}
Shifting the field $\varphi(x) \to \varphi_c(x) + \varphi(x)$, we can write the imaginary part of the partition function in the one-loop approximation as
\begin{equation}
\begin{aligned}
\text{Im} Z & \simeq \frac{e^{A/g}}{i} \int \mathscr D  \varphi \, e^{-\frac{1}{2} \int \! d^d x_1 d^d x_2 \,  \varphi(x_1) M(x_1,x_2) \varphi(x_2)} \\
& = \frac{e^{A/g}}{i} \left( \text{det} \, M M_0^{-1} \right)^{-1/2}\,,
\end{aligned}
\end{equation}
where $M_0$ is the operator with kernel
\begin{equation}
M_0(x_1,x_2) \equiv  \left[ -\Delta_{x_1} + 1 \right] \delta(x_1-x_2) \,,
\end{equation}
that is, the second functional derivative of the free action. However one has to notice that the operator $M$ has $d$ zero modes given by $\partial_\mu \Phi_c(x)$ as can be seen by simple differentiation of the equation of motion (\ref{EOM2}). This is of no surprise: the instanton solution, which depends the $d$ parameters $x_{0 \mu}$, breaks explicitly the translational invariance of the action. We will see in the next section how to treat these zero modes using the \emph{collective coordinates method}, which properly integrates over the parameters $x_{0 \mu}$. As anticipated, note that the operator $MM_0^{-1}$ has one, and only one, negative eigenvalue; for this reason the square root of the determinant is imaginary and the final result is real, as it should be. 

\subsection{Collective Coordinates Method}

In full generality we will call 
\begin{equation}
\chi_\mu(x) = \partial_\mu \Phi_c(x)\,, \qquad \mu = 1, \dots, d 
\end{equation}
the $d$ zero modes of $M$. In order to integrate explicitly over the parameters $x_{0 \mu}$, with $\mu = 1, \dots , d$, we introduce the simple identity
\begin{equation}
\begin{aligned}
& 1 = \prod_{\mu=1}^{d}  \int \frac{d x_{0 \mu}}{\sqrt{2 \pi}}  \left[ \int d^d x \, \chi_\mu(x) \partial_\mu \varphi(x+ x_{0 \mu} \hat \mu) \right] \\
& \exp \left\{ - \frac{1}{2} \left[  \int d^d x \, \chi_\mu(x) \left( \varphi(x + x_{0 \mu} \hat \mu) - \varphi_c(x) \right) \right]^2 \right\}
\end{aligned}
\end{equation}
in our functional integral
\begin{equation}
\begin{aligned}
\text{Im} Z & = \frac{1}{2i} \! \int\!\frac{d^d x_0}{(2\pi)^{d/2}} \int \! \mathscr D \varphi \, e^{-S_\chi(\varphi; \, x_0)} .
\end{aligned}
\end{equation}
We have defined
\begin{equation}
\begin{aligned}
& S_{\chi}(\varphi; x_0) \equiv S(\varphi) - F_\chi(\varphi, x_0) + \frac{1}{2} \sum_{\mu=1}^{d} \left[ \int \! d^d x \, \right. \\
& \qquad \qquad \quad \Biggl. \chi_\mu(x-x_{0 \mu} \hat \mu) \left( \varphi(x) - \varphi_c(x- x_{0 \mu} \hat \mu ) \right) \Biggr]^2 \,,
\end{aligned}
\end{equation}
and the Jacobian
\begin{equation}
\begin{aligned}
& F_\chi(\varphi; x_0)   \equiv \sum_{\mu=1}^{d} \ln \int \! d^d x \, \chi_\mu(x - x_{0 \mu} \, \hat \mu ) \, \partial_\mu \varphi(x) \,.
\end{aligned}
\end{equation}
Note that the new action is composed of a term translationally invariant, $S(\varphi)$, and of a term which breaks explicitly the symmetry. However, performing the change of variables $\varphi \to \varphi(x- x_{0\mu} \hat \mu)$, $x \to x - x_{0 \mu} \hat \mu$, it is easy to see that the dependence of $x_0$ disappears. This means that we get a factor proportional to the volume $V$
\begin{equation}
\label{ImZStartingPoint}
\begin{aligned}
\text{Im} Z & = \frac{ V}{2i \left(2\pi\right)^{d/2}} \! \int \! \mathscr D \varphi \, e^{-S_\chi(\varphi)}\,,
\end{aligned}
\end{equation}
where we will call
\begin{equation}
\begin{aligned}
& S_\chi (\varphi) \equiv S_\chi(\varphi, 0) \,, \\
& F_\chi (\varphi) \equiv F_\chi(\varphi, 0) \,.
\end{aligned}
\end{equation}
Now we proceed expanding around the instanton saddle point. The additional terms in $S_\chi(\varphi)$, does not change the saddle point of the old action $S(\varphi)$. In fact, the equation of motion
\begin{equation}
\begin{aligned}
0 & = \frac{\delta S_\chi}{\delta \varphi (x_1)}  = \frac{\delta S }{\delta \varphi(x_1)} +\\
& + \sum_{\mu=1}^{d} \left[ \int d^d x \, \chi_\mu(x) \left( \varphi(x) - \varphi_c(x) \right) \right] \chi_\mu (x_1)  \,,
\end{aligned}
\end{equation}
is satisfied for the usual saddle point $\varphi(x) = \varphi_c(x)$. We have not included the derivative of $F_\chi(\varphi)$ because it is of order $\sqrt{-g}$, so it will be relevant only at the two-loop order. When we compute the second functional derivative
\begin{equation}
\label{SecondDerivativeAction}
\left. \frac{\delta^2 S_\chi}{\delta \varphi(x_1) \delta \varphi(x_2)} \right|_{\varphi = \varphi_c } = M(x_1, x_2) 
+ \sum_{\mu=1}^{d} \chi_\mu(x_1) \chi_\mu(x_2)
\end{equation}
we get an additional term that is proportional to a sum over the projectors onto zero modes of $S(\varphi )$; therefore this will shift only the zero eigenvalues.

The leading order expansion of the Jacobian $F_{\chi}(\varphi)$ will produce a factor of the type
\begin{equation}
e^{F_\chi(\varphi_c)} = \frac{1}{\left( -g \right)^{d/2}} \prod_{\mu=1}^{d} \alpha_\mu\,.
\end{equation}
where we have denoted with $\alpha_\mu$ the quantity
\begin{equation}
\label{alphamuDef}
\alpha_\mu = \|\partial_\mu \Phi_c\|^2  = \int d^d x \left( \partial_\mu \Phi_c\right)^2\,,
\end{equation}
Using the spherical symmetry of the solution we can see that $\alpha_\mu$ does not really depend on $\mu$
\begin{equation}
\label{alphamu}
\begin{aligned} 
\alpha_\mu & = \frac{1}{d} \int d^d x \, \left( \nabla \Phi_c(x) \right)^2 = A \,,
\end{aligned}
\end{equation}
so that
\begin{equation}
\label{InstantonNorm}
\begin{aligned}
e^{F_\chi(\varphi_c)} = \frac{  A^{d} }{(-g)^{d/2}} = \left[-\frac{1}{g} \left(\frac{I_4}{4}\right)^2 \right]^{d/2} \,.
\end{aligned}
\end{equation}
Since our mass operator is that of equation (\ref{SecondDerivativeAction}), we have to evaluate
\begin{equation}
\det B = \det \left[ M+\sum_\mu \ket{\chi_\mu} \bra{\chi_\mu} \right]/ \det M_0
\end{equation}
where 
\begin{equation}
\begin{aligned}
B & \equiv {1 \over \sqrt{M_0}}  \left[ M+\sum_\mu \ket{\chi_\mu} \bra{\chi_\mu} \right]{1 \over \sqrt{M_0}} \\
&= I-Q+\sum_\mu {1 \over \sqrt{M_0}}\ket{\chi_\mu} \bra{\chi_\mu} {1 \over \sqrt{M_0}} \,.
\end{aligned}
\end{equation}
$Q$ is the Hermitian operator
\begin{equation}
Q \equiv 3 {1 \over \sqrt{M_0}} \Phi_c^2(x) {1 \over \sqrt{M_0}} \,,
\end{equation}
so it has orthogonal eigenfunctions $\overline{\psi_i}$ with eigenvalues $\frac{1}{\lambda_i}$
\begin{equation}
\label{QSpectrumEq}
Q \, \ket{ \overline{\psi}_i } = \frac{1}{\lambda_i} \ket{ \overline{\psi}_i} \,.
\end{equation} 
Numerically one can compute $\lambda_i$ integrating 
\begin{equation}
\label{EigenvalueEquation}
(-\Delta + 1) \, \psi_i = 3 \lambda_i \Phi_c^2(x) \, \psi_i
\end{equation} 
where we define $\ket{\psi_i}$
\begin{equation}
\label{PsiBar}
\ket{\psi_i} \equiv M_0^{-1/2}  \ket{\overline{\psi}_i } \implies  \bra{\psi_i} M_0 \ket{\psi_i} = 1 \,.
\end{equation}
The last equality follows from the fact that $\ket{\overline{\psi}_i }$ are normalized. As a consequence the last term in the definition of $B$ is diagonal in the base of $Q$ because $Q$ has eigenvectors with eigenvalue equal to $1$ corresponding to $\chi$; they are given by
\begin{equation}
\ket{\overline{\chi}_\mu } \equiv \frac{M_0^{1/2} \ket{\chi_\mu}}{ \sqrt{\bra{\chi_\mu} M_0 \ket{\chi_\mu} } } \,.
\end{equation}
To write the last term in the definition of $B$ in terms of the eigenvectors of $Q$, we evaluate the matrix element
\begin{equation}
\label{ElementMatrix}
\bra{\overline{\chi}_{\mu'}} \frac{1}{\sqrt{M_0}} \ket{\chi_\mu} \bra{\chi_\mu} \frac{1}{\sqrt{M_0}} \ket{\overline{\chi}_{\nu }} = \delta_{\mu \mu'} \, \delta_{\mu \nu} \frac{\braket{\chi_\mu | \chi_\mu }^2 }{\bra{\chi_\mu} M_0 \ket{\chi_\mu} } \,.
\end{equation}
Using again the spherical symmetry of the instanton solution, we can write respectively the numerator and denominator of (\ref{ElementMatrix}) as
\begin{subequations}
	\begin{equation}
	\braket{\chi_\mu | \chi_\mu }^2 = \left[\frac{1}{d} \int d^d x \, \left( \nabla \Phi_c \right)^2 \right]^2 = A^2 = \left( \frac{I_4}{4} \right)^2 \,,
	\end{equation}
	\begin{equation}
	\label{chiM0chi}
	\bra{\chi_\mu} M_0 \ket{\chi_\mu} = \frac{3}{d} \int \! d^d x \left( \nabla \Phi_c \right)^2 \Phi_c^2  = \frac{I_6-I_4}{d} \,.
	\end{equation}
\end{subequations}
Note that the $\ket{ \psi_i }$ corresponding to $\lambda_i  = 1$ are not equal to $\ket{ \chi_\mu}$.
Using these results and expanding on the eigenvalues of $Q$, given in equation (\ref{QSpectrumEq}) we obtain the essential relationship
\begin{equation}
\label{B_EigenExpansion}
B= \sum_{i, \lambda_i \neq 1} \left( 1- {1 \over \lambda_i}\right) \ket{\overline{\psi}_i } \bra{\overline{\psi}_i } + {   d ( I_4/4)^2 \over (I_6-I_4) } \sum_{i, \lambda_i=1} \ket{\overline{\psi}_i } \bra{\overline{\psi}_i } 
\end{equation}
From the above expression we can express the determinant of $B$ as
\begin{equation}
\label{Determinant}
\det{B} = \left[ {  d (I_4/4)^2 \over (I_6-I_4) } \right]^d \Omega^\bot    \,,
\end{equation}
where we have defined the determinant of the fluctuation operator $B$ projected onto the space orthogonal to the zero modes 
\begin{equation}
\label{OrthogonalDeterminant}
\Omega^\bot =  \prod_{i, \lambda_i \neq 1} \left( 1- {1 \over \lambda_i}\right) \,.
\end{equation}
In this last expression it is understood that a degenerate eigenvalue will appear in the product a number of times equal to its degeneracy. 
In (\ref{Determinant}) we have taken into account the fact that there are $d$ eigenvalues with $\lambda_i=1$ that were obtained from the zero modes of $M$.
In the final result for $\mathrm{Im} Z$ we have to multiply by $\left( \det B \right)^{-1/2}$ and the term $(I_4/4)^{-d}$ cancels with a similar term coming from the Jacobian of the collective coordinates transformation as in equation (\ref{InstantonNorm}). At the leading order, the final effect of the collective coordinates method is to produce a factor $(-g)^{-d/2}$. 

In 
\ref{AppendixA} we derive result (\ref{Determinant}) using time independent perturbation theory. 

\subsection{Determinant in one dimension}

Only in one dimension the product (\ref{OrthogonalDeterminant}) is convergent. Indeed, since the instanton solution is (\ref{Instanton1d}), the eigenvalue equation (\ref{EigenvalueEquation}) can be easily solved because the potential is of the Bargmann type \cite{Brezin&Guillou&ZJ}. The eigenvalues are
\begin{equation}
\lambda_i = \frac{i(i+1)}{6}\,, \qquad i = 1, 2, \dots
\end{equation}  
so that the determinant of the operator $B$ restricted to the subspace orthogonal to zero modes is
\begin{equation}
\Omega^\bot \equiv \prod_{i=1, \, i\ne 2}^{\infty} \left( 1- \frac{1}{\lambda_i} \right) = -2 \prod_{i=3}^{\infty} \left( 1- \frac{6}{i(i+1)} \right)  \,.
\end{equation}
Performing the product we get 
\begin{equation}
\begin{aligned}
&\prod_{i=3}^{\infty} \left( 1- \frac{6}{i(i+1)} \right) = \prod_{i=3}^{\infty} \left( \frac{i-2}{i} \, \frac{i+3}{i+1} \right) \\
&  = \lim\limits_{n\to \infty}  \frac{2}{n(n-1)} \frac{(n+3)(n+2)}{20} = \frac{1}{10}
\end{aligned}
\end{equation}
so that
\begin{equation}
\Omega^\bot = -\frac{1}{5}.
\end{equation}

\subsection{One-Loop Renormalization}

Collecting all factors, the imaginary part of the partition function is
\begin{equation}
\text{Im} Z = V  \frac{e^{I_4/4g}}{\left(-g\right)^{d/2}} \left[\frac{I_6-I_4}{2\pi d} \right]^{d/2} \left[ -\Omega^\bot \right]^{-1/2} e^{\frac{1}{2g} \delta m^2_{(1)} I_2}
\end{equation}
where the last term comes from the first order expansion of the mass counterterm which is
\begin{equation}
\delta m^2 \simeq \delta m^2_{(1)} = -3 g G_0(0) \,,
\end{equation}
where
\begin{equation}
G_0(x) = \int \frac{d^d p }{(2 \pi)^d} \frac{e^{ipx}}{p^2+1} \,.
\end{equation}
$G_0(0)$ is divergent in $d = 2$ and $3$. Now we have to verify that the mass counterterm renormalizes the divergence coming from the $\Omega^{\bot}$. Indeed one can write $\Omega^\bot$ in terms of traces of the operator $Q^n$ as
\begin{equation}
\label{OmegaBot}
\Omega^\bot = \exp \left\{- \sum_{n=1}^{\infty} \frac{1}{n} \left( \text{Tr} \, Q^n - d\right)  \right\} \,.
\end{equation}
Representing the $\Phi_c^2(y)$ insertion with a wiggly line 
one can clearly see that $\text{Tr} \, Q^n$ is associated with the one-loop diagrams with $2n$ external $\Phi_c^2$ legs. Therefore only the $n=1$ term in (\ref{OmegaBot}) is divergent in 2 and 3 dimensions and it is given by
\begin{equation}
\text{Tr} \, Q = \sum_{k=0}^{\infty} \frac{1}{\lambda_k} = \!\!\!
\parbox{14mm}{
	\begin{fmffile}{f.2}
	\begin{fmfgraph*}(35,30)
	\fmfleft{i1}
	\fmfright{o1}
	\fmftop{t0,t1,t2}
	\fmfbottom{b0,b1,b2}
	\fmf{phantom}{i1,v,o1}
	\fmf{plain, left, tension=0.2}{v,b1,v}
	\fmffreeze
	\fmf{wiggly}{v,t1}
	\end{fmfgraph*}
	\end{fmffile}
}  \!\!\!\! 
= 3 G_2(0) \, I_2 \,,
\end{equation}
This divergence is canceled exactly by the mass counterterm. We can therefore define a renormalized determinant as
\begin{equation}
\label{RenormalizedDeterminant}
\Omega^\bot_R \equiv \Omega^\bot e^{3G_0(0) I_2 } = e^d \prod_{i, \lambda_i \ne 1} \left( 1- \frac{1}{\lambda_i} \right) e^{\frac{1}{\lambda_i}}
\end{equation}
The final result is
\begin{equation}
\text{Im} Z = V  \frac{e^{I_4/4g}}{\left(-g\right)^{d/2}} \left[\frac{I_6-I_4}{2\pi d} \right]^{\frac{d}{2}} \left[ -\Omega^\bot_R \right]^{-\frac{1}{2}} \left[1+O(g)\right]
\end{equation}

The determinant $\Omega^\bot_R$ has been evaluated in dimension 2 and 3 by \cite{Parisi}, using the \emph{Thomas-Fermi approximation}, which gives an additional information coming from the asymptotic behavior of the large $\lambda_k$'s. We review this procedure, in 
\ref{AppendixB}. In Table \ref{tab:Determinant} we report our numerical value of $\Omega^\bot_R$ in $d=2$, $3$ and the analytical value of $\Omega^\bot$ in one dimension (there is no need of renormalization in $d=1$). Note that the numerical value of $\Omega^\bot_R$ in $d=2$ we present is different from that of \cite{Parisi} probably due to a typo.

\begin{table}
	\begin{center}
		\begin{tabular}[b]{c|ccc}
			\hline
			$d$ & $1$ & $2$ & $3$  \\  
			\hline
			$\Omega_R^\bot$ & --  & $-88.7823$  & $-10.5459$  \\  
			$\Omega^\bot$ & $- \frac{1}{5}$  & --  & --  \\  
			\hline
		\end{tabular}
	\end{center}
	\caption{ Numerical estimate of $\Omega_R^\bot$ in dimension 2 and 3 using the Thomas-Fermi approximation. We also report the analytical value of $\Omega^\bot$ in one dimension.}\label{tab:Determinant}
\end{table}

\subsection{The Free Energy}

Using the dispersion relation (\ref{DispersionRelation}), we obtain the one-loop large order behavior of the partition function
\begin{equation}
Z_k \simeq V \, C \left(-A\right)^{-k} \,\Gamma\left( k + \frac{d}{2}\right)  \left[1 + O\left(\frac{1}{k}\right) \right] \,,
\end{equation}
where 
\begin{equation}
C =  \frac{1}{\pi} \left[\frac{2 \left(I_6-I_4\right)}{\pi d \, I_4} \right]^{d/2} \left[ -\Omega^\bot_R \right]^{-1/2} \,.
\end{equation}
In order to evaluate the large order behavior of the free energy
\begin{equation}
F(g) = - \lim\limits_{V \to \infty} \frac{1}{V} \ln Z(g) = \sum_{k=0}^{\infty} F_k \, g^k\,,
\end{equation}
one can use
\begin{equation}
\label{GroundStateEnergy}
F_k = - \frac{Z_k}{V} \left[ 1 - Z_1 \frac{Z_{k-1}}{Z_{k}}  + O \left( \frac{1}{k^2} \right)\right] \,.
\end{equation}
Since $Z_k$ grows like $k!$, the term $Z_{k-1}/Z_k$ is of order $1/k$, and it can be neglected at this order. We obtain
\begin{equation}
F_k = -C \left(-A\right)^{-k} \,\Gamma\left( k + \frac{d}{2}\right)  \left[1 + O\left(\frac{1}{k}\right) \right] \,.
\end{equation}

\section{Two-Loop Order Expansions}
\label{Section3}

We will call the coefficient that contributes to the order $g$ in the expansion of the imaginary part of $Z$ as $\xi$ 
\begin{equation}
\text{Im} Z =  V  \frac{e^{I_4/4g}}{\left(-g\right)^{d/2}} \left[\frac{I_6-I_4}{2\pi d} \right]^{\frac{d}{2}} \left[ -\Omega^\bot_R \right]^{-\frac{1}{2}} \left[1+\xi g \right] \,.
\end{equation}
Using again the dispersion relation (\ref{DispersionRelation}) the $1/k$ correction to the large order behavior of $Z_k$ will be written in terms of $\xi$ as
\begin{equation}
\label{ZkTwoLoopOrder}
Z_k \simeq  V \, C \left(-A\right)^{-k} \,\Gamma\left( k + \frac{d}{2}\right)  \left[1 - \frac{\xi I_4}{4} \frac{1}{k} + O \left( \frac{1}{k^2} \right) \right] .
\end{equation}

Starting from the expression in equation (\ref{ImZStartingPoint}), $\xi$ will have contribution from
\begin{itemize}
	\item $F_\chi(\varphi)$, i.e. the Jacobian of the collective coordinate change of variable;
	\item the third and fourth functional derivative of $S(\varphi)$ evaluated on the instanton; 
	\item the mass counterterm $\delta m^2$.
\end{itemize}

Note that the projector onto the zero modes $\chi_\mu$ which is present in $S_\chi(\varphi)$ and its second functional derivative (\ref{SecondDerivativeAction}) does not give any contribution to the third and fourth derivative of the action. The only non-trivial term coming from the collective coordinates that gives a contribution is the Jacobian $F_\chi(\varphi)$. This is also what happens in the one-dimensional case \cite{Auberson,Collins&Soper, Jen&Zinn}, where the Jacobian contributes only with a one-leg vertex of order $\sqrt{-g}$. We will see that, in $d=2$ and $d=3$, there will be another additional vertex of order $g$. In the following subsections, we will analyze with care all these contributions mentioned above. We will recover the one-dimensional result and extend it to field theory.

\subsection{Collective Coordinates Jacobian}
The expansion of $F_\chi(\varphi)$ is
\begin{multline} 
F_\chi(\varphi)  = \sum_{\mu=1}^{d} \ln \left[ \frac{\alpha_\mu}{\sqrt{-g}} +\int \! d^d x \; \partial_\mu \Phi_c(x) \, \partial_\mu \varphi(x) \right] \\
= \sum_{\mu=1}^{d} \left[ \ln \frac{\alpha_\mu}{\sqrt{-g}} + \ln \left(  1 - \frac{\sqrt{-g}}{\alpha_\mu d}  \int \! d^d x \; \Delta \Phi_c(x) \, \varphi(x)  \right) \right]
\end{multline}
The first term in the right hand side is the one-loop contribution. Expanding the logarithm and using (\ref{alphamu}), we have  
\begin{equation}
\label{JacobianSecondOrder}
\begin{aligned} 
F_\chi(\varphi) & = \sum_{\mu=1}^{d}  \ln  \alpha_\mu - \frac{\sqrt{-g}}{ A }  \int \! d^d x \; \Delta \Phi_c(x) \, \varphi(x) \\
&  +\frac{1}{ 2} \frac{g }{A^2 d} \left[  \int \! d^d x \; \Delta \Phi_c(x) \, \varphi(x) \right]^2
\end{aligned}
\end{equation}
We obtain two types of vertices, one of order $\sqrt{-g}$ and the other of order $g$
\begin{equation}   
\begin{aligned}
\sigma & = - \frac{\sqrt{-g}}{ A}   &\longrightarrow  \;\;
\parbox{16mm}{
	\begin{fmffile}{f.Jacobian}
	\begin{fmfgraph*}(40,25)
	\fmfv{decor.shape=circle, decor.size=5}{i}
	\fmfleft{i}
	\fmfright{o}
	\fmf{phantom}{v,o}
	\fmf{plain}{i,v}
	\end{fmfgraph*}
	\end{fmffile}
}   \\
\tau & = \frac{g}{ 2 A^2 d} &\longrightarrow  \;\;
\parbox{16mm}{
	\begin{fmffile}{f.Jacobian2}
	\begin{fmfgraph*}(20,25)
	\fmfv{decor.shape=circle, decor.size=5}{i,o}
	\fmfleft{i}
	\fmfright{o}
	\fmftop{v1,v2}
	\fmf{plain}{i,v1}
	\fmf{plain}{o,v2}
	\end{fmfgraph*}
	\end{fmffile}
}  
\end{aligned}
\end{equation} 
respectively with one and two external legs.

\subsection{Action}
The third and fourth derivative of the action are
\begin{equation*}
\begin{aligned} 
&  -\left.\frac{\delta^3 S}{\delta \varphi (x_1) \delta \varphi (x_2) \delta \varphi (x_3)}  \right|_{\varphi=\varphi_c} \\
& = 6\sqrt{-g} \, \Phi_c(x) \, \delta(x_1 -x_2) \, \delta(x_1 -x_3)  \\
& -\left. \frac{\delta^4 S}{\delta \varphi (x_1) \delta \varphi (x_2) \delta \varphi (x_3) \delta \varphi(x_4)}\right|_{\varphi=\varphi_c}  \\
& = -6g  \, \delta(x_1 -x_2) \, \delta(x_1 -x_3) \, \delta(x_1-x_4)  
\end{aligned}
\end{equation*}
We keep the factors 1/3! and 1/4! in order to maintain the usual symmetry factors in the diagrams. One can define three and four-legs vertices as
\begin{equation}
\begin{aligned}
&\lambda = 6\sqrt{-g}   \quad & \longrightarrow &
&\parbox{20mm}{
	\begin{fmffile}{f.3}
	\begin{fmfgraph*}(40,25)
	\fmfv{decor.shape=triacross, decor.size=20}{v}
	\fmfleft{i2,i1}
	\fmfright{o2,o1}
	\fmfkeep{triacross}
	\fmf{phantom}{i1,v,o2}
	\fmf{phantom}{i2,v,o1}
	\end{fmfgraph*}
	\end{fmffile}
} \\
&  \alpha = -6g    & \longrightarrow  &
&\parbox{20mm}{
	\begin{fmffile}{f.4}
	\begin{fmfgraph*}(40,25)
	\fmfv{decor.shape=cross, decor.size=20}{v}
	\fmfleft{i2,i1}
	\fmfright{o2,o1}
	\fmfkeep{cross}
	\fmf{phantom}{i1,v,o2}
	\fmf{phantom}{i2,v,o1}
	\end{fmfgraph*}
	\end{fmffile}
}  
\end{aligned}
\end{equation}

\subsection{Counterterms}
At the two loop order we can write the mass counterterm as
\begin{equation}
\delta m^2 = \delta m^2_{(1)} + \delta m^2_{(2)} = -3g G_0(0) + 6 g^2 K(0) \,,
\end{equation}
where
\begin{equation}
\label{K(p)}
K(p) = \int \!\frac{d^d k_1}{(2 \pi)^d} \frac{d^d k_2}{(2 \pi)^d}  \frac{1}{k_1^2 + 1} \,  \frac{1}{k_2^2 + 1} \,  \frac{1}{(p-k_1-k_2)^2 + 1} \,.
\end{equation}
When $p=0$ this is logarithmically divergent in $d=3$ and convergent in $d=2$. Expanding the action counterterm around the instanton solution, we have 
\begin{equation}
\begin{aligned} 
- &\frac{\delta m^2}{2} \int d^d x \, \varphi^2(x) = -\frac{3}{2}G_0(0) I_2 + 3gK(0) I_2 \\
& +\gamma \int d^d x \, \Phi_c(x) \, \varphi(x) +\frac{\delta}{2} \int d^d x \, \varphi^2(x) \,,
\end{aligned} 
\end{equation}
where $\gamma$ and $\delta$ represents two additional vertices, with one and two external legs respectively. We denote them by a little cross 
\begin{equation}
\begin{aligned}
&  \gamma = -3 \sqrt{-g} \, G_0(0)  & \longrightarrow &
&\parbox{16mm}{
	\begin{fmffile}{f.5}
	\begin{fmfgraph*}(40,25)
	\fmfv{decor.shape=cross, decor.size=5}{i}
	\fmfleft{i}
	\fmfright{o}
	\fmf{phantom}{v,o}
	\fmf{plain}{i,v}
	\fmfkeep{OneLeg}
	\end{fmfgraph*}
	\end{fmffile}
}  \\
&  \delta = 3g \, G_0(0)  & \longrightarrow  &
&\parbox{16mm}{
	\begin{fmffile}{f.6}
	\begin{fmfgraph*}(40,25)
	\fmfv{decor.shape=cross, decor.size=5}{v}
	\fmfleft{i}
	\fmfright{o}
	\fmf{plain}{i,v,o}
	\fmfkeep{twoLegs}
	\end{fmfgraph*}
	\end{fmffile}
} 
\end{aligned}
\end{equation} 
Obviously these vertices are not present in dimension 1, since renormalization is not needed.

\section{Diagrams}
\label{Section4}

In the previous section we have obtained, expanding the action around the instanton, a theory with cubic and quartic vertices respectively of order $\sqrt{-g}$ and $g$. This is of no surprise because the same happens in the perturbative expansion of the $\varphi^4$ theory below the critical temperature \cite{ParisiLibro}, where one has a potential with two degenerate, non-vanishing minima. In our case, since we have $g<0$, the potential is the same apart of a minus sign factor. The really big difference is that the saddle point is not only non-vanishing, but is also space-time dependent. The propagator of our theory is  
\begin{equation}
L(x,y) = \bra{x} \left[  M + \sum_{\mu=1}^{d} \chi_\mu \chi_\mu \right]^{-1} \ket{y} \,.
\end{equation}
Combining all vertices to the order $g$ we obtain the diagrams
\begin{equation*}
\begin{aligned}
& \parbox{14mm}{
	\begin{fmffile}{f.7}
	\begin{fmfgraph*}(35,20)
	\fmfleft{i}
	\fmfright{o}
	\fmf{plain,tension=0.3}{v1,v2}
	\fmf{plain, left, tension=0.1}{v1,i,v1}
	\fmf{plain, left, tension=0.1}{v2,o,v2}
	\fmfkeep{manubrio}
	\end{fmfgraph*}
	\end{fmffile}
} = \int \! d^d x \, d^d y \, \Phi_c(x)  L(x,x)  L(x,y)  L(y,y)  \Phi_c(y) \\
& \parbox{14mm}{
	\begin{fmffile}{f.8}
	\begin{fmfgraph*}(35,20)
	\fmfleft{i}
	\fmfright{o}
	\fmf{phantom,tension=5}{i,v1}
	\fmf{phantom,tension=5}{v2,o}
	\fmf{plain,left,tension=0.5}{v1,v2,v1}
	\fmf{plain}{v1,v2}
	\fmfkeep{sunset}
	\end{fmfgraph*}
	\end{fmffile}
} = \int \!d^d x \, d^d y \, \Phi_c(x) L^3(x,y)  \Phi_c(y) \\
&  \parbox{14mm}{
	\begin{fmffile}{f.9}
	\begin{fmfgraph*}(35,20)
	\fmfleft{i}
	\fmfright{o}
	\fmftop{t0}
	\fmfbottom{b0}
	\fmf{phantom}{t0,v,b0}
	\fmffreeze
	\fmf{plain, left, tension=0.5}{v,o,v}
	\fmf{plain, left, tension=0.5}{v,i,v}
	\fmfkeep{eight}
	\end{fmfgraph*}
	\end{fmffile}
} = \int \! d^d x \, L^2(x,x)\\
&\parbox{14mm}{
	\begin{fmffile}{f.10}
	\begin{fmfgraph*}(35,20)
	\fmfleft{i}
	\fmfright{o}
	\fmfkeep{tadpole}
	\fmf{plain}{i,v}
	\fmf{plain, left, tension=0.5}{v,o,v}
	\fmfv{decor.shape=cross, decor.size=5}{i}
	\end{fmfgraph*}
	\end{fmffile}
}  =  \int \! d^d x \, d^d y \, \Phi_c(x) L(x,y) L(y,y) \Phi_c(y)\\
&\parbox{14mm}{
	\begin{fmffile}{f.12}
	\begin{fmfgraph*}(35,20)
	\fmfleft{i}
	\fmfright{o}
	\fmfkeep{blobx}
	\fmf{phantom}{i,v}
	\fmf{plain, left, tension=0.5}{v,o,v}
	\fmfv{decor.shape=cross, decor.size=5}{v}
	\end{fmfgraph*}
	\end{fmffile}
} = \int \! d^d x \, L(x,x) \\
&\parbox{14mm}{
	\begin{fmffile}{f.11}
	\begin{fmfgraph*}(30,20)
	\fmfleft{i}
	\fmfright{o}
	\fmfkeep{plainx}
	\fmf{plain}{i,o}
	\fmfv{decor.shape=cross, decor.size=5}{i,o}
	\end{fmfgraph*}
	\end{fmffile}
}  = \int \! d^d x \, d^d y \, \Phi_c(x) L(x,y) \Phi_c(y) \\
&\parbox{14mm}{
	\begin{fmffile}{f.13}
	\begin{fmfgraph*}(35,20)
	\fmfleft{i}
	\fmfright{o}
	\fmfkeep{tadpoleJac}
	\fmf{plain}{i,v}
	\fmf{plain, left, tension=0.5}{v,o,v}
	\fmfv{decor.shape=circle, decor.size=5}{i}
	\end{fmfgraph*}
	\end{fmffile}
}  =  \int \! d^d x \, d^d y \, \Delta \Phi_c(x) L(x,y) L(y,y) \Phi_c(y)\\
& \parbox{14mm}{
	\begin{fmffile}{f.14}
	\begin{fmfgraph*}(30,20)
	\fmfleft{i}
	\fmfright{o}
	\fmfkeep{plainxJac}
	\fmf{plain}{i,o}
	\fmfv{decor.shape=circle, decor.size=5}{i}
	\fmfv{decor.shape=cross, decor.size=5}{o}
	\end{fmfgraph*}
	\end{fmffile}
}  = \int \! d^d x \, d^d y \, \Delta \Phi_c(x) L(x,y) \Phi_c(y) \\
& \parbox{14mm}{
	\begin{fmffile}{f.15}
	\begin{fmfgraph*}(30,20)
	\fmfleft{i}
	\fmfright{o}
	\fmfkeep{plainxJacTwo}
	\fmf{plain}{i,o}
	\fmfv{decor.shape=circle, decor.size=5}{i}
	\fmfv{decor.shape=circle, decor.size=5}{o}
	\end{fmfgraph*}
	\end{fmffile}
}  = \int \! d^d x \, d^d y \, \Delta \Phi_c(x) L(x,y) \Delta \Phi_c(y) \\
\end{aligned}
\end{equation*}
which have multiplicity respectively 
\begin{subequations}
	\begin{align}
	\label{manubrio}
	\manubrio{14mm} & \longrightarrow  \, \frac{\lambda^2}{8} = - \frac{9}{2} g \,; \\
	\label{sunset}
	\sunset{14mm} &\longrightarrow  \, \frac{\lambda^2}{3! \, 2} =  - 3 g \,; \\
	\label{eight}
	\eight{14mm} &\longrightarrow \, \frac{\alpha}{8} = - \frac{3}{4} g \,; \\
	\label{tadpole}
	\tadpole{14mm} &\longrightarrow \, \frac{\gamma \lambda}{2} = 9 g \, G_0(0) \,; \\
	\label{blobx}
	\blobx{14mm} &\longrightarrow   \, \frac{\delta}{2} = \frac{3}{2} g \, G_0(0) \,; \\
	\label{plainx}
	\plainx{14mm} &\longrightarrow \, \frac{\gamma^2}{2} = -\frac{9}{2} g \, G_0^2(0) \,; \\
	\label{tadpoleJac}
	\tadpoleJac{14mm} &\longrightarrow \, \frac{\sigma \lambda}{2} =  \frac{3g}{A } \,; \\
	\label{plainxJac}
	\plainxJac{14mm} &\longrightarrow \, \sigma \gamma  = -\frac{3g}{A } G_0(0) \,; \\
	\label{plainxJacTwo}
	\plainxJacTwo{14mm} &\longrightarrow \, \frac{\sigma^2}{2} + \tau  = - \frac{g(d-1)}{ 2 A^2 d} \,.
	\end{align}
\end{subequations}
Note that to this order we do not obtain disconnected diagrams. The two-loop correction to the imaginary part of the partition function $\xi$, is expressed in terms of diagrams as
\begin{equation}
\label{zeta}
\begin{aligned}
& \xi =  - \frac{9}{2} \left[ \manubrio{13mm} - 2G_0(0) \tadpole{13mm} + \right. \\
& \left.  + G_0^2(0) \plainx{13mm} \right] - 3 \left[\!\!\!\!\!\sunset{10mm}  - I_2 K(0) \right] \\
& - \frac{3}{4} \left[ \eight{13mm} -2G_0(0) \!\!\!\!\!\!\!\! \blobx{13mm} \right] \\
& +\frac{3}{A} \left[\tadpoleJac{14mm} \! -G_0(0) \, \plainxJac{14mm}  \! \right] \\
& - \frac{(d-1)}{ 2 A d} \plainxJacTwo{14mm}
\end{aligned}
\end{equation}

\subsection{The One-Dimensional Case}
When $d=1$, only diagrams (\ref{manubrio}), (\ref{sunset}), (\ref{eight}) and (\ref{tadpoleJac}) contribute, consistently with the result of \cite{Auberson, Collins&Soper, Jen&Zinn}. Note that diagram (\ref{plainxJacTwo}), cancels because of the multiplicity factor proportional to $d-1$. Following \cite{Auberson} in this case $L(x,y)$ can be calculated exactly and it is given by
\begin{multline}
\label{1dPropagator}
L(x,y) = \pm \frac{1}{2} \left[ u(x) \, v(y) - v(x) \,u(y) \right] + \\
\frac{\sinh x \sinh y}{4} \left[ \frac{1}{\cosh^2 x } +\frac{1}{\cosh^2 y } - \frac{1}{\cosh^2 x \cosh^2y} \right]
\end{multline} 
where
\begin{subequations}
	\begin{align}
	u(x) & = \frac{\sinh x}{\cosh^2 x} \\
	v(x) & =  - \frac{1}{2} \cosh x + \frac{3}{2 \cosh x} - \frac{3 x}{2}\frac{\sinh x}{\cosh^2 x} \,.
	\end{align}
\end{subequations}
In Eq. (\ref{1dPropagator}) we take the plus sign if $x<y$ and the minus sign otherwise. Using the explicit expression of the propagator one can evaluate analytically the diagrams. Their value is
\begin{subequations}
	\begin{align}
	\manubrio{14mm} &= \frac{13}{168}\,; \\
	\sunset{14mm} & = -\frac{71}{2016}\,; \\
	\eight{14mm} & = \frac{V}{4} - \frac{319}{420}\,; \\
	\tadpoleJac{14mm} &= \frac{53}{180} \,. 
	\end{align}
\end{subequations}
Using that $A = I_4/4 = 4/3$ in one dimension, the two loop correction $\xi$ will be
\begin{equation}
\xi = \frac{95}{96} - \frac{3}{16} V \,.
\end{equation}

\subsection{Renormalization} 



The propagator $L$ can be written in terms of $B$ as
\begin{equation}
L = \frac{1}{\sqrt{M_0}} \frac{1}{B} \frac{1}{\sqrt{M_0}}
\end{equation}
The inverse of $B$ can be easily written using its diagonal base (see equation (\ref{B_EigenExpansion})), so that we can write
\begin{multline}
L = \sum_{i, \lambda_i \neq 1} \left( 1- {1 \over \lambda_i}\right)^{-1} \frac{1}{\sqrt{M_0}}  \ket{\overline{\psi}_i } \bra{\overline{\psi}_i } \frac{1}{\sqrt{M_0}}  \\
+ {   (I_6-I_4)  \over d ( I_4/4)^2 } \sum_{i, \lambda_i=1} \frac{1}{\sqrt{M_0}}  \ket{\overline{\psi}_i } \bra{\overline{\psi}_i } \frac{1}{\sqrt{M_0}} \,.
\end{multline}
Using equation (\ref{PsiBar}) and (\ref{chiM0chi}), we have
\begin{multline}
L = \sum_{i, \lambda_i \neq 1} \left( 1- {1 \over \lambda_i}\right)^{-1}  \ket{ \psi_i } \bra{\psi_i }   \\
+ {  1 \over ( I_4/4)^2 } \sum_{\mu}   \ket{ \chi_\mu } \bra{ \chi_\mu } 
\end{multline}
The above result can be plugged into the diagrams directly in $d=1$. In dimension 2 and 3, instead, one has to take care of renormalization and a different expression must be used. The trick is to extract the free propagator $\frac{1}{M_0}$ from the expression of $L$; this surely will permit us to isolate the infinite contribution of each diagram and to check that these divergences will be canceled by counterterms. We write
\begin{equation}
\label{PropagatorSplit}
L = \frac{1}{M_0} + H \,,
\end{equation}
where
\begin{multline}
\label{H(x,y)}
H 
=  \sum_{i, \lambda_i \neq 1} \frac{1}{\lambda_i-1}  \ket{ \psi_i } \bra{\psi_i }  \\
+\left( {   (I_6-I_4) \over d ( I_4/4)^2 } - 1 \right) \sum_{i, \lambda_i = 1}   \ket{ \psi_i } \bra{\psi_i } \,.  
\end{multline}  
Note that every time we have a propagator $L$ with equal arguments, we will get a divergence of the type $G_0(0)$ coming from the first term in equation (\ref{PropagatorSplit}). Using the decomposition (\ref{PropagatorSplit}), we can write the diagrams in this way
\begin{subequations}
	\begin{gather}
	\begin{multlined}
	\label{LastDivergence} 
	\eight{13mm} -2G_0(0) \!\!\!\!\!\!\!\! \blobx{13mm}  = - V G_0^2(0) \\
	+ \int \! d^d x \, H^2(x,x) \,,
	\end{multlined}
	\\
	\begin{multlined}
	\label{Manubrio}
	\manubrio{13mm} - 2G_0(0) \tadpole{13mm} + G_0^2(0) \plainx{13mm}  \\
	\!\!\!= \int \! d^d x \, d^d y \; \Phi_c(x) H(x,x) L(x,y) H(y,y) \Phi_c(y)\,,  
	\end{multlined}
	\\
	\begin{multlined}
	\label{Tadpole}
	\tadpoleJac{14mm} \! -G_0(0) \, \plainxJac{14mm}  \! \\
	= \int \! d^d x \, d^d y \, \Delta \Phi_c(x) L(x,y) H(y,y) \Phi_c(y)  
	\end{multlined}
	\end{gather}
\end{subequations}
We expect that the preceding integral expressions are finite. The only diagram which is not completely renormalized is that of equation (\ref{LastDivergence}). We will see later that the additional divergence $  V G_0^2(0)$ is a feature of the partition function; once we compute the free energy of the system this infinity will be canceled.

Substituting $L(x,y) \to G_0(x-y)$ in (\ref{eight}), and passing to Fourier space, we get
\begin{equation}
\sunset{12mm} \!\! = \int \! \frac{d^d p}{(2 \pi)^d} \, \Phi_c^2(p) \, K(p)
\end{equation}
When $k_1$ and $k_2$ in (\ref{K(p)}) are large, this expression reduces to
\begin{equation}
\sunset{12mm} \!\! \simeq I_2 K(0) \,,
\end{equation}
which cancels exactly with the corresponding counterterm. With these considerations in mind we obtain
\begin{equation}
\label{Sunset}
\begin{aligned}
\!\!\!\!\! \sunset{12mm} \!\!\!  & - I_2 K(0)  = S + R
\end{aligned}
\end{equation}
where
\begin{subequations}
	\begin{gather}
	\begin{multlined}
	\label{S}
	S = \int \! d^d x \, d^d y \; \Phi_c(x) \left[ 3 G_0^2(x-y) \, H(x,y) \right. \\ 
	\left. + 3 G_0(x-y) \, H^2(x,y) + H^3(x,y)\right] \Phi_c(y) \,,
	\end{multlined}
	\\
	\begin{multlined}
	\label{R}
	R =  \int \!\frac{d^d p}{(2 \pi)^d} \, \Phi_c^2(p) 
	\left[ K(p) - K(0) \right] \,.
	\end{multlined}
	\end{gather}
\end{subequations}
We expect that also these two quantities are finite. The difference $K(p) - K(0)$ can be written as \cite{ParisiLibro}
\begin{equation}
\begin{aligned}
& K(p) - K(0) = \frac{\Gamma(3-d)}{(4 \pi)^d} \int_{0}^{1} \! \frac{dx \, dy \, dz}{(xy + zx + yz)^{d/2}}  \\
&\left[ \left( \frac{xyz}{xy + zx + yz} p^2 + 1 \right)^{d-3} - 1 \right] \delta(x+y+z-1) \,.
\end{aligned}
\end{equation}
This expression can be directly used in $d=2$, since the $K(0)$ counterterm do not have divergences. In $d=3$, the $K(0)$ subtraction cancels the pole of $\Gamma(3-d)$. Explicitly
\begin{equation}
\begin{aligned}
& K(p) - K(0) = -\frac{1}{(4 \pi)^3} \int_{0}^{1} \! \frac{dx \, dy \, dz}{(xy + zx + yz)^{3/2}}  \\ 
& \ln \left( \frac{xyz}{xy + zx + yz} p^2 + 1 \right) \delta(x+y+z-1) \,.
\end{aligned}
\end{equation}

\subsection{Free Energy}

As anticipated, we expect that, when we compute the free energy, the last divergence isolated in equation (\ref{LastDivergence}) will be canceled in $d=2, 3$. Starting from (\ref{ZkTwoLoopOrder}) we compute the ratio
\begin{equation}
\frac{Z_{k-1}}{Z_{k}} = - \frac{I_4}{4k} \,.
\end{equation}
and 
\begin{equation}
Z_1 = \frac{3}{4} V G_0^2(0) \,,
\end{equation}
Plugging these two results into identity (\ref{GroundStateEnergy}), we have
\begin{equation}
\begin{aligned}
F_k & = -C \left(-A\right)^{-k} \,\Gamma\left( k + \frac{d}{2}\right) \\
& \times \left[ 1 - \left( \xi - \frac{3}{4} V G_0^2(0) \right) \frac{I_4}{4 k}\right].
\end{aligned}
\end{equation}
Therefore the infinite contribution coming from $Z_1$ completely renormalizes the last divergence. 

When $d=1$, instead, we have to use
\begin{equation}
Z_1 = - \frac{3}{16} V \,,
\end{equation}
since the free propagator is $G_0(x) = e^{-x}/2$. The two loop correction to the free energy is
\begin{equation}
- \left( \xi - \frac{3}{16} V \right) \frac{I_4}{4} = - \frac{95}{72} \,.
\end{equation}
This final result can be matched with that of \cite{Auberson, Collins&Soper, Jen&Zinn}.

\section{Numerical Results}
\label{Section5}

In order to lighten the notation we rewrite $H$ as a sum over all eigenvalues
\begin{equation}
\begin{aligned} 
H =  \sum_{i} \frac{1}{\lambda_i-1}  \ket{ \psi_i } \bra{\psi_i } \,;
\end{aligned}
\end{equation}
in which it is assumed that the eigenvalue equal to one must be replaced by 
\begin{equation}
1 + \left( {   (I_6-I_4) \over d ( I_4/4)^2 } - 1 \right)^{-1} \,,
\end{equation}

to be consistent with equation (\ref{H(x,y)}).

\subsection{Case d=2}

In $d=2$ the eigenfunctions are decomposed as
\begin{equation}
\psi_i(x) =  R_{nm}(r_x) \frac{e^{im \phi_x}}{ \sqrt{2 \pi} } \,,
\end{equation}
$H(x,y)$ can be written as a function of $r_x$, $r_y$ and $\theta = \phi_y-\phi_x$, i.e. the angle between $x$ and $y$
\begin{equation}
H(r_x,r_y,\theta) = \frac{1}{2 \pi} \sum_{n = 1 }^{\infty} \sum_{m=-\infty}^{+\infty} \frac{e^{i m \theta } }{\lambda_{n m} -1 } R_{nm}(r_x) R_{nm}(r_y) \,.
\end{equation}
The multiplicity can be highlighted using only the sum over $m \ge 0$
\begin{equation}
\begin{aligned}
H(r_x,r_y, \theta) & = \frac{1}{2 \pi} \sum_{n = 1 }^{\infty} \sum_{m=0}^{+\infty} \frac{1}{\lambda_{n m} -1 } R_{nm}(r_x)  \\
& \times \left[ 2 \cos\left( m \theta \right) - \delta_{m, 0}\right] R_{nm}(r_y) \,.
\end{aligned}
\end{equation}
For $x=y$ this expression no longer depends on the angle $\theta$
\begin{equation}
\begin{aligned}
H(r_x) \equiv H(r_x,r_x, 0) & = \frac{1}{2 \pi} \sum_{n = 1 }^{\infty} \sum_{m=0}^{+\infty} \frac{2 - \delta_{m, 0}}{\lambda_{n m} -1 } R^2_{nm}(r_x)  \\
\end{aligned}
\end{equation}

\subsection{Case d=3}

In $d=3$ the eigenfunctions are decomposed in this way
\begin{equation}
\psi_i(x) =  R_{nl}(r_x) \, Y_{lm}(\theta_x, \phi_x) \,,
\end{equation}
Using the identity
\begin{equation}
\sum_{m=-l}^{l} Y^*_{lm}(\theta_x, \phi_x) \, Y_{lm}(\theta_y, \phi_y) = \frac{2l+1}{4 \pi} P_l(\cos \theta) \,,
\end{equation}
where $P_l$ is the Legendre polynomial of degree $l$ and $\theta$ is the angle between $x$ and $y$, $H(x,y)$ can be written as
\begin{equation}
H(r_x,r_y,\theta) = \frac{1}{4 \pi} \sum_{n = 1}^{\infty} \sum_{l=0}^{n-1} \frac{2l+1}{\lambda_{n l} -1 } P_l(\cos \theta) R_{nl}(r_x) R_{nl}(r_y) 
\end{equation}
that is a function of $r_x$, $r_y$ and $\theta$. 
For $x=y$ the dependence on $\theta$ drops
\begin{equation}
H(r_x) \equiv H(r_x,r_x,0) = \frac{1}{4 \pi} \sum_{n = 1}^{\infty} \sum_{l=0}^{n-1} \frac{2l+1}{\lambda_{n l} -1 } R^2_{nl}(r_x) \,.
\end{equation}

\subsection{Final Expressions of diagrams}
\setlength{\extrarowheight}{.1em}
\begin{table}
	\begin{center}
		\begin{tabular}[b]{@{}  c|cc @{} }
			\hline
			& $d=2$ & $d=3$  \\  
			\hline
			(\ref{BlobFinal}) & $0.0555027(6)$ & $0.058015(6)$\\
			(\ref{ManubrioFinal}) & $-0.11569(2)$ & $0.38075(8)$\\  
			(\ref{TadpoleFinal}) & $-0.20829(2) $ & $10.0620(5)$\\  
			(\ref{plainxJacTwoFinal}) & $-0.0130032(1)$ & $-14.170027(1)$\\   
			$S$ & $-0.0875141(4)$ & $-1.1295(3)$\\    
			$R$ & $-7.19815\cdot 10^{-3}$ & $-5.45801\cdot 10^{-3}$\\  
			\hline
		\end{tabular}
	\end{center}
	\caption{Numerical estimate of Diagrams. The errors come from the extrapolation to an infinite number of eigenfunctions. }
	\label{tab:Diagrams}
\end{table}

\begin{table}
	\begin{center}
		\begin{tabular}[b]{c|cc}
			\hline
			& $d=2$ & $d=3$  \\  
			\hline
			$\xi - \frac{3}{4}  V G_0^2(0)$ & $0.6553(2)$ & $3.4988(9) $ \\
			\hline
		\end{tabular}
	\end{center}
	\caption{Numerical estimate of $\xi - \frac{3}{4}  V G_0^2(0)$. The errors come from the extrapolation to an infinite number of eigenfunctions. }
	\label{tab:xi}
\end{table}

Since the dipendence on $r_x$, $r_y$ and $\theta$ is present both in 2 and 3-dimensional expression of $H$ and in the free propagator 
\begin{equation}
G_0(r_x, r_y, \theta) \equiv 
G_0\left(\sqrt{ r_x^2 + r_y^2 -2 r_x r_y \cos \theta } \right) 
\end{equation} 
one can write the 2d-dimensional integrals in diagrams (\ref{Manubrio}),  (\ref{Tadpole}) and (\ref{Sunset}) as
\begin{multline}
\int \! d^d x \, d^d y \; f(r_x,r_y,\theta) = \Omega_{d-1} \,\Omega_{d-2} \int \! d r_x \, d r_y \\
\int_0^\pi \! d \theta \, r_x^{d-1} r_y^{d-1} \left(\sin \theta \right)^{d-2} f(r_x,r_y,\theta) \,,
\end{multline}
where $f(r_x,r_y,\theta)$ is the specific function to integrate and $\Omega_{d-1}$ is the surface area of the unit radius $d$-sphere
\begin{equation}
\Omega_{d-1} = \frac{2 \pi^{d/2} }{ \Gamma \left( d/2 \right)} \,.
\end{equation}
Diagram (\ref{LastDivergence}) reads
\begin{equation}
\label{BlobFinal}
\int d^d x \,H^2(x,x) = \Omega_{d-1} \int d r_x \, r_x^{d-1} \, H^2(r_x) \,.
\end{equation}
In diagrams (\ref{Manubrio}),  (\ref{Tadpole}) and (\ref{plainxJacTwo}), one further simplification can be made. Since there is only one $H(r_x, r_y, \theta)$ term, one can perform exactly the integration over $\theta$. In $d=3$ one can use the properties of Legendre Polynomials
\begin{equation}
\int_{-1}^{1} d \theta \, P_l(\theta) = 2 \int_{0}^{1} d \theta \, P_{2l} (\theta) = 2 \delta_{l, 0} \,,
\end{equation}
instead in $d=2$ the integral is simply
\begin{equation}
\int_{0}^{\pi} d \theta \cos  \left( m \theta \right) = \pi \delta_{l, 0} \,.
\end{equation}
The previously mentioned diagrams can be written as
\begin{subequations}
	\begin{gather}
	\begin{multlined}
	\label{ManubrioFinal}
	(\ref{Manubrio}) = \Omega_{d-1} \Omega_{d-2} \int \! d r_x \, d r_y \, r_x^{d-1}  r_y^{d-1} \\ 
	\times \Phi_c(r_x) H(r_x) L(r_x,r_y)  H(r_y) \Phi_c(r_y) 
	\end{multlined}
	\\
	\begin{multlined}
	\label{TadpoleFinal}
	(\ref{Tadpole}) = \Omega_{d-1} \Omega_{d-2} \int \! d r_x \, d r_y \, r_x^{d-1}  r_y^{d-1} \\
	\times \Delta \Phi_c(r_x) L(r_x,r_y)  H(r_y) \Phi_c(r_y) 
	\end{multlined}
	\\
	\begin{multlined}
	\label{plainxJacTwoFinal}
	(\ref{plainxJacTwo}) = \Omega_{d-1} \Omega_{d-2} \int \! d r_x \, d r_y \, r_x^{d-1}  r_y^{d-1} \\ 
	\times \Delta \Phi_c(r_x) L(r_x,r_y)  \Delta\Phi_c(r_y) 
	\end{multlined}
	\end{gather}
\end{subequations}
where $L(r_x,r_y)$ is
\begin{equation}
L(r_x,r_y) = \int_{0}^{\pi} \! d \theta \left(\sin \theta \right)^{d-2} G_0(r_x,r_y, \theta) + H(r_x, r_y)
\end{equation}
and $H(r_x, r_y)$ is given by
\begin{equation}
H(r_x, r_y) = \left\{
\begin{aligned}
&\frac{1}{2} \sum_{nm} \frac{ \delta_{m, 0} }{\lambda_{n m} -1 } R_{nm}(r_x) R_{nm}(r_y)\,, \\
& \frac{1}{2\pi} \sum_{nl} \frac{ \delta_{l, 0} }{\lambda_{n l} -1 } R_{nl}(r_x) R_{nl}(r_y)\,,
\end{aligned}
\right.
\end{equation}
respectively in $d=2$ and $d=3$.

In Table \ref{tab:Diagrams} we report the numerical estimate of diagrams and of quantity $R$ defined in (\ref{R}). In Table \ref{tab:xi} the numerical value of the correction $\xi- \frac{3}{4} V G_0^2(0)$ is given. All the errors come from a linear extrapolation to an infinite number of eigenfunction.

\section{Conclusions} \label{Conclusions} 

In this work we have discussed, using the Lipatov approach, how to evaluate the large order behavior of a perturbative series. We have taken into consideration the simple case of the partition function of the $\varphi^4$ theory. We have computed the $1/k$ correction to the perturbative coefficient of the partition function $Z_k$ and of the free energy $F_k$. In dimension 1, our results coincide with that of Bender and Wu \cite{Bender&Wu1} using a WKB approach and of \cite{Auberson, Collins&Soper, Jen&Zinn} using Lipatov method. When $d=2$ or 3, we have shown that there are the same diagrams present in dimension 1, plus one that comes from the Jacobian of the collective coordinates method. All these quantities have been evaluated numerically.

Knowing the large order behavior has proven to be important in field theory since one can transform the original divergent perturbation series in a convergent one. The best way to do that has been achieved using the Borel summation method combined with a conformal transformation. In particular, the one-loop evaluation of large orders gives the value of the singularity of the Borel transform closest to the origin; this information is essential in order to use the conformal mapping itself. We stress that these techniques proved to be very versatile since they can also be applied to random systems. We mention here the percolation problem (obtained from a cubic $O(n)$ theory in the limit $n\to 0$) \cite{WallaceZia,Houghton,McKane,Carlucci}, the random diluted Ising model \cite{BrayMoore,McKaneIsing,MartinMayor} and the $\varepsilon$ expansion in spin glasses in the paramagnetic phase \cite{Moore}. Note also that the imaginary part of the partition function is connected with the decay rate of metastable states \cite{Langer67}. 

In the case of pure $\varphi^4$ theory, an estimate of the two-loop term of the large perturbative series is important to improve the precision of physical quantities like critical exponents \cite{GuidaZinn}. Since the perturbative series is known to the seventh order in $d=3$ \cite{MurrayNickel}, the evaluation of the eighth order is surely much harder than the computation of the two-loop term in the high order expansion. However this computation is not completely trivial both at the analytical and at the numerical level. This work can be considered as a preliminary investigation that clarifies these conceptual problems by analyzing the simplest case of the partition function. We leave for future work the investigation of the more physically interesting case of the $2M$-point Green function which are directly connected to experimentally measurable quantities such as critical exponents.

\section*{Acknowledgements}

This work was supported by grant No. 454949 from the Simons Foundation.

\appendix
\section{Obtaining (\ref{Determinant}) by Perturbation Theory}
\label{AppendixA}
\begin{figure}[!ht]
	\centering
	\includegraphics[width=\columnwidth,height=1.1\columnwidth]{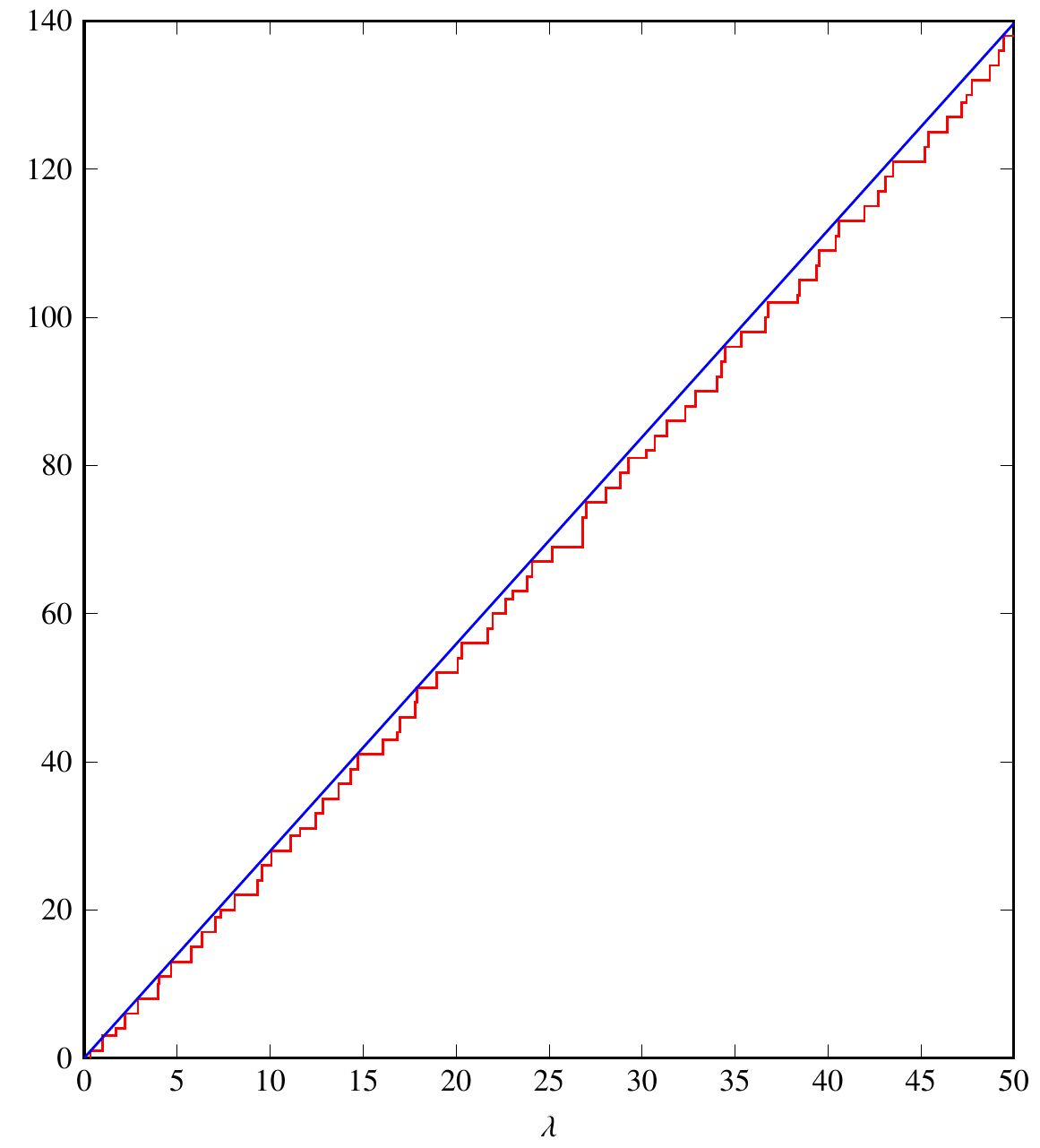}
	\caption{$N(\lambda)$ as a function of $\lambda$ in $d=2$: the red line is the numerical result; the blu line is $N_{\text{as}}(\lambda)$, the asymptotic Thomas-Fermi estimate (see equation (\ref{Nas})). } 
	\label{Fig:N2}
\end{figure}

Using the definition of $\alpha_\mu$ in (\ref{alphamuDef}), the determinant will be
\begin{equation}
\det \left[ \left( M + \sum_{\mu=1}^{d} \ket{\chi_\mu} \bra{\chi_\mu} \right) M_0^{-1} \right] = \Omega \, \prod_{\mu=1}^{d} \alpha_\mu = \Omega \left( \frac{I_4}{4}\right)^d\,,
\end{equation}
where
\begin{equation}
\label{Omega}
\begin{aligned}
\Omega & \equiv \lim\limits_{\varepsilon \to 0} \frac{1}{\varepsilon^d} \det\left[ \left( M + \varepsilon \right) \left( M_0 + \varepsilon \right)^{-1} \right] \\
& = \lim\limits_{\varepsilon \to 0} \frac{\det \left( 1- \Xi(\varepsilon) \right)}{\varepsilon^d}
\end{aligned}
\end{equation}
and 
\begin{equation}
\label{XiDef}
\Xi (\varepsilon) = 3 \left(M_0+\varepsilon\right)^{-1} \Phi_c^2(x) = \frac{3}{- \Delta + 1 + \varepsilon} \Phi_c^2(x) \,.
\end{equation}
$\Xi(\varepsilon)$ is \emph{not} an Hermitian operator, but it has real and positive spectrum since it is equivalent to an operator 
\begin{equation}
\label{QDef}
Q(\varepsilon) = 3 \left(M_0 + \varepsilon \right)^{-1/2} \Phi_c^2(x) \left(M_0 + \varepsilon \right)^{-1/2} \,,
\end{equation}
via 
\begin{equation}
\Xi(\varepsilon) = \left( M_0 + \varepsilon \right)^{-1/2} Q(\varepsilon) \left(M_0 + \varepsilon \right)^{1/2}
\end{equation}
which is Hermitian and positive \cite{ZJ, Parisi}. We will call the eigenvalues of $\Xi(\varepsilon)$ by $1/\lambda_i(\varepsilon)$, $\, i=0, 1, \dots$; for simplicity we will denote them by $1/\lambda_i$ when $\varepsilon=0$. They can be obtained by a numerical solution of the equation
\begin{equation}
\label{SpectrumPerturbedEquation}
\left(-\Delta + 1 + \varepsilon \right) \psi_i = 3 \lambda_i(\varepsilon) \, \Phi_c^2(x) \, \psi_i \,.
\end{equation} 
In the limit $\varepsilon \to 0$ we know that the ground state and the first ($d$-times degenerate) excited state solution of this equation are
\begin{equation}
\begin{aligned}
& \lambda_0 = \frac{1}{3} \,, \quad \psi_0(x) = \Phi_c(x)  \,; \\
& \lambda_1 = 1 \,, \quad \psi_1(x) = \partial_\mu \Phi_c(x) \,.
\end{aligned}
\end{equation}
$\lambda_0$ will give the only negative contribution to the determinant. Since $\left( 1 - 1/\lambda_1(\varepsilon) \right)$ is of order $\varepsilon$ when $\varepsilon \ne 0$ but small, we can write $\Omega$ defined in equation (\ref{Omega}), in terms of $\Omega^\bot$ as
\begin{subequations}
	\begin{equation}
	\Omega = \Omega^\bot \lim\limits_{\varepsilon \to 0} \left( \frac{1- \frac{1}{\lambda_1(\varepsilon)} }{\varepsilon} \right)^d
	\end{equation}
	\begin{equation}
	\Omega^\bot =  \prod_{i, \lambda_i \ne 1} \left( 1 - \frac{1}{\lambda_i} \right)  \,.
	\end{equation}
\end{subequations}
(\ref{SpectrumPerturbedEquation}) is an equation of the \gatto type with potential $-3 \lambda_i(\varepsilon) \Phi_c^2(x)$ and energy $-1-\varepsilon$. Since we want to evaluate the first order expansion of $\lambda_1(\varepsilon)$ in $\varepsilon$, we can use first order perturbation theory. The perturbing Hamiltonian will be
\begin{equation}
\begin{aligned}
\mathscr{H}_p & = - 3 \Phi_c^2(x) (\lambda_1(\varepsilon)-1) \\
& = - 3 \Phi_c^2(x) \left(1- \frac{1}{\lambda_1(\varepsilon)} \right) + O(\varepsilon)
\end{aligned}
\end{equation}
and the corresponding energy shift is $-\varepsilon$. The unperturbed eigenstates $\partial_\mu \Phi_c(x)$ are $d$ times degenerate. However, the perturbing Hamiltonian is diagonal in the space of unperturbed eigenstates $\partial_\mu \Phi_c(x)$, with all equal elements on the diagonal. In fact, using the invariance under space rotations of $\Phi_c(x)$ we have
\begin{equation}
\bra{\partial_\mu \Phi_c } \mathscr{H}_p\ket{\partial_\nu \Phi_c } = \frac{\delta_{\mu \nu}}{d} \int d^d x \, \Phi_c^2(x) \left(  \nabla \Phi_c(x) \right)^2 \,.
\end{equation}
We conclude that $\lambda_1(\varepsilon)$ remains $d$ times degenerate at first order in $\varepsilon$. Therefore we can then use simple time independent, non-degenerate perturbation theory; using again the spherical symmetry of the instanton solution $\Phi_c(x)$, the energy shift is
\begin{align}
-\epsilon & = \frac{\bra{\partial_\mu \Phi_c } \mathscr{H}_p\ket{\partial_\mu \Phi_c }}{ \braket{\partial_\mu \Phi_c | \partial_\mu \Phi_c} } \\
&= -3 \left(1-\frac{1}{\lambda_1(\varepsilon)} \right) \, \frac{\int d^d x \; \Phi_c^2(x) \left( \nabla \Phi_c(x) \right)^2 }{\int d^d x \, \left( \nabla \Phi_c(x) \right)^2} \,.
\label{PerturbationTheory}
\end{align}
Using an integration by parts
\begin{equation}
\begin{aligned}
\int d^d x \; \Phi_c^2(x) \left( \nabla \Phi_c(x) \right)^2 & = -2 \int d^d x \; \Phi_c^2(x) \left( \nabla \Phi_c(x) \right)^2 \\
& - \int d^d x \; \Phi_c^3(x) \left( \Delta \Phi_c(x) \right) 
\end{aligned}
\end{equation}
and the equation of motion (\ref{EOM2}), we have that the numerator of (\ref{PerturbationTheory}) is
\begin{equation}
\int d^d x \; \Phi_c^2(x) \left( \nabla \Phi_c(x) \right)^2 = \frac{1}{3} \left( I_6 - I_4 \right) \,.
\end{equation}
The denominator is simply
\begin{equation}
\int d^d x \, \left( \nabla \Phi_c(x) \right)^2 = I_4-I_2 = \frac{1}{4} d  I_4 \,.
\end{equation}
Inserting into (\ref{PerturbationTheory}) we get 
\begin{equation}
\lim\limits_{\varepsilon \to 0 } \frac{1- \frac{1}{\lambda_1(\varepsilon)}}{\varepsilon} = \frac{\frac{1}{4} d I_4}{I_6-I_4} \,.
\end{equation}
Collecting all factors we get the same result obtained in (\ref{Determinant}).

\section{The Thomas-Fermi Approximation} \label{AppendixB}

\begin{figure}[!ht]
	\centering
	\includegraphics[width=\columnwidth,height=1.1\columnwidth]{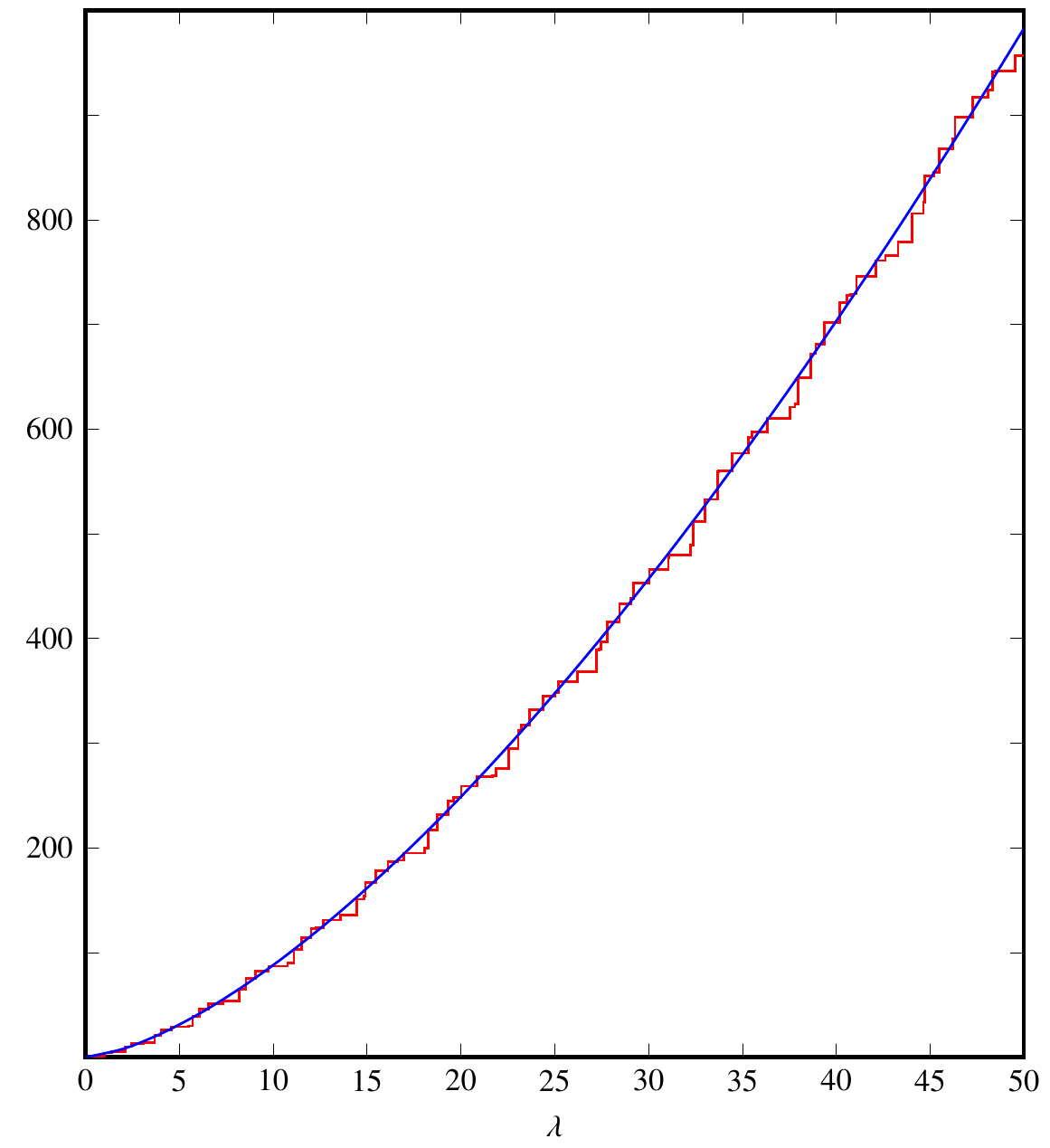}
	\caption{$N(\lambda)$ as a function of $\lambda$ in $d=3$: the red line is the numerical result; the blu line is $N_{\text{as}}(\lambda)$, the asymptotic Thomas-Fermi estimate (see equation (\ref{Nas})). } 
	\label{Fig:N3}
\end{figure}
Let us consider the generic Hamiltonian
\begin{equation}
\mathscr H(p,x) = \sum_{\mu=1}^{d} p_\mu^2 - \lambda V(x) \,, \qquad V(x)>0\,,
\end{equation}
where the potential $V(x)$ goes to zero at infinity. Let us call $N(\lambda)$ the number of eigenvalues smaller then $\lambda$. $N(\lambda)$ will have a jump all the time there is an eigenvalue and the height of the jump corresponds to the degeneracy of eigenvalue itself. The \emph{Thomas-Fermi approximation} gives us the behaviour of $N(\lambda)$ for large $\lambda$ \cite{ParisiLibro}
\begin{equation}
\lim\limits_{\lambda \to \infty} N(\lambda) = N_{\text{as}}(\lambda) = \frac{\Omega_{d-1}}{d} \int \! \frac{d^d x}{\left( 2 \pi \right)^d} \left( \lambda V(x) \right)^{d/2} \,.
\end{equation}
Note that $N_{\text{as}}(\lambda)$ is a continuous function of its argument. Differentiating with respect to $\lambda$ one gets the asymptotic spectral density
\begin{equation}
\rho_{\text{as}}(\lambda) = \frac{d N_{\text{as}}}{d \lambda} = \frac{\Omega_{d-1}}{2} \lambda^{d/2-1}\int \! \frac{d^d x}{\left( 2 \pi \right)^d} \left( V(x) \right)^{d/2} \,. 
\end{equation}
In our case $V(x) = 3 \Phi_c^2(x)$, so that
\begin{align}
\label{Nas}
&  N_{\text{as}}(\lambda) = \frac{3^{d/2}}{ \left( 2 \pi \right)^d} \frac{\Omega_{d-1}}{d} I_d \lambda^{d/2} \,, \\
& \rho_{\text{as}}(\lambda) = \frac{3^{d/2} \Omega_{d-1}}{2 \left( 2 \pi \right)^d }  I_d \lambda^{d/2-1}\,.
\end{align}
In Fig. \ref{Fig:N2} and \ref{Fig:N3} one can see a comparison between the numerical $N(\lambda)$ and the asymptotic Thomas-Fermi estimate (\ref{Nas}), in dimension 2 and 3 respectively. 
This information can be of great utility in order to numerically evaluate the determinant of fluctuation operator orthogonal to zero eigenmodes \cite{Parisi}. Suppose we perform the product in (\ref{RenormalizedDeterminant}) until the eigenvalue $\overline{\lambda}$. We can then use 
\begin{equation}
\begin{aligned}
& \prod_{i \ne 1} \left(1-\frac{1}{\lambda_i} \right) e^{\frac{1}{\lambda_i}} \simeq \prod_{\lambda_k < \overline{\lambda} } \left(1-\frac{1}{\lambda_i} \right) e^{\frac{1}{\lambda_i}}  \\
& \exp\left\{  \int_{\overline{\lambda}}^{\infty} d^d \lambda \, \rho_{\text{as}}(\lambda) \left[ \ln \left(1 - \frac{1}{\lambda}\right) + \frac{1}{\lambda} \right] \right\} \\ 
& \simeq e^{ - C_d \overline{\lambda}^{d/2-2} } \prod_{\lambda_k < \overline{\lambda} } \left(1-\frac{1}{\lambda_i} \right) e^{\frac{1}{\lambda_i}}  
\end{aligned}
\end{equation}
where
\begin{equation}
C_d = \frac{3^{d/2} \Omega_{d-1}}{2 \left( 2 \pi \right)^d }  \frac{I_d}{4-d} \,.
\end{equation}
The convergence of $\Omega_R^\bot$ with the Thomas-Fermi factor is relatively faster.

\end{document}